\documentclass[preprint,amsmath,showpacs]{revtex4-1}
\usepackage[latin9]{inputenc}
\usepackage{color}
\usepackage{amsmath}
\usepackage{amssymb}
\usepackage{graphicx}
\usepackage{epstopdf}
\usepackage{esint}

\makeatletter
\@ifundefined{textcolor}{}
{%
 \definecolor{BLACK}{gray}{0}
 \definecolor{WHITE}{gray}{1}
 \definecolor{RED}{rgb}{1,0,0}
 \definecolor{GREEN}{rgb}{0,1,0}
 \definecolor{BLUE}{rgb}{0,0,1}
 \definecolor{CYAN}{cmyk}{1,0,0,0}
 \definecolor{MAGENTA}{cmyk}{0,1,0,0}
 \definecolor{YELLOW}{cmyk}{0,0,1,0}
}

\makeatother

\begin{document}
\pagestyle{plain}

\title{Probing the coupling of heavy dark matter to nucleons by detecting
neutrino signature from the Earth's core}

\author{Guey-Lin Lin, Yen-Hsun Lin and Fei-Fan Lee}

\affiliation{Institute of Physics, National Chiao Tung University, Hsinchu 30010,
Taiwan}
\date{\today}
\begin{abstract}
We argue that the detection of neutrino signature from the Earth's core
can effectively probe the coupling of heavy dark matter
($m_{\chi}>10^{4}$ GeV) to nucleons. We first note that direct searches
for dark matter (DM) in such a mass range provide much less stringent
constraint than the constraint provided by such searches for $m_{\chi}\sim 100$ GeV. 
Furthermore the energies of neutrinos arising from DM
annihilation inside the Sun cannot exceed a few TeVs at the Sun surface
due to the attenuation effect. Therefore the sensitivity to the heavy
DM coupling is lost. Finally, the detection of neutrino signature
from galactic halo can only probe DM annihilation cross sections.
We present neutrino event
rates in IceCube and KM3NeT arising from the neutrino flux produced
by annihilation of Earth-captured DM heavier than $10^{4}$ GeV.
The IceCube and KM3NeT sensitivities to spin independent DM-proton
scattering cross section $\sigma_{\chi p}$ in this mass range are presented for both  isospin symmetric and 
isospin violating cases. 
\end{abstract}

\maketitle

\section{Introduction}

Evidences for the dark matter (DM) are provided by many astrophysical observations, although the nature of DM is yet to be uncovered. The most popular candidates for DM are weak interacting massive particles (WIMP), which we shall assume in this work. DM can be detected either directly or indirectly with the former observing the nucleus recoil as DM interacts
with the target nuclei in the detector and the latter detecting final state particles resulting from DM
annihilation or decays. The direct detection is possible because that 
the dark matter particles constantly bombard the Earth as the Earth sweeps through the local 
halos. 
Sensitivities to $\sigma_{\chi p}$ from DM direct searches are low for large $m_{\chi}$. Given a fixed DM mass density $\rho_{\rm DM}$ in the solar system, the number density of DM particles is inversely proportional to $m_{\chi}$. Furthermore, the nuclear form factor suppression is more severe for DM-nucleus scattering for large $m_{\chi}$. For a review of direct detection, see~\cite{Cushman:2013zza}. 

In this work, we propose to probe the coupling of heavy DM to nucleons by indirect approach with neutrinos.
We note that the flux of DM induced neutrinos from galactic halo is only sensitive
to $\left\langle \sigma\upsilon\right\rangle $. Furthermore, the
energies of neutrinos from the Sun can not exceed a few TeVs due to
severe energy attenuation during the propagation inside the Sun. Hence, for $m_{\chi}> 10^4$ GeV, we turn to the possibility of probing such DM with the search of neutrino signature from the Earth's core.

It is important to note that, for $E_{\nu}\gtrsim 100$ TeV, all flavors of neutrinos interact with nucleons inside the Earth with a total cross section $\sigma\varpropto E^{0.5}$~\cite{Gandhi:1996}. Charged-current (CC) and neutral-current (NC) neutrino-nucleon interactions occur in the ratio 0.71:0.29 and the resulting lepton carries about 75 $\%$ of the initial neutrino energy in both cases~\cite{Gandhi:1996}. During CC interaction,
initial $\nu_{e}$ and $\nu_{\mu}$ will disappear and the resulting $e$ and $\mu$ will be brought to rest due to their electromagnetic energy losses. Thus high-energy $\nu_{e}$ and $\nu_{\mu}$ are absorbed by the Earth. However the situation is very different for $\nu_{\tau}$ \cite{Ritz:1988,Halzen:1998be}, because except for very high energies ($\gtrsim 10^{6}$ TeV), the tau lepton decay length is less than its range, so that the tau lepton decays in flight without significant energy loss. In every branch of tau decays, $\nu_{\tau}$ is produced. In this regeneration process $\nu_{\tau}\rightarrow\tau\rightarrow\nu_{\tau}$, the regenerated $\nu_{\tau}$ carries about 1/3 of the initial $\nu_{\tau}$ energy \cite{Dutta:2000,Gaisser:1992}. Those $\nu_{\tau}$ arriving at the detector site can be identified through shower events. We further note that 18$\%$ of the tau decays are $\tau\rightarrow\nu_{\tau}\mu\overline{\nu}_{\mu}$ and another 18$\%$ are $\tau\rightarrow\nu_{\tau}e\overline{\nu}_{e}$. These secondary anti-neutrinos $(\overline{\nu}_{e}, \overline{\nu}_{\mu})$ carry roughly 1/6 of the initial $\nu_{\tau}$ energy. The secondary 
$\overline{\nu}_{\mu}$ flux is detectable as muon track events or hadronic shower events. Similarly, the
secondary $\overline{\nu}_{e}$ flux is also detectable as shower events~\cite{Beacom:2002}. In summary, as tau neutrinos propagate through the Earth, the regenerated tau neutrinos by prompt tau decays can produce relatively large fluxes of secondary $\overline{\nu}_{e}$ and $\overline{\nu}_{\mu}$ and hence greatly enhance the detectability of the initial $\nu_{\tau}$. Therefore we study the neutrino signature from DM annihilation channels $\chi\chi\rightarrow\tau^{+}\tau^{-}, \, W^+W^-$, and $\nu\bar{\nu}$ from the Earth's core.

The status of IceCube search for neutrinos coming from DM annihilation in the Earth's core has been reported~\cite{Kunnen:2013}. The earlier IceCube data on the search for astrophysical muon neutrinos was used to constrain the cross section of DM annihilation  
$\chi\chi\to \nu\bar{\nu}$ in the Earth's core~\cite{Albuquerque:2011ma} for $m_{\chi}$ in the favored range of PAMELA and Fermi experiments~\cite{Adriani:2011xv,Ackermann:2010ij}. The sensitivity of IceCube-DeepCore detector to various DM annihilation channels in the Earth's core for low mass DM has also been studied in Ref.~\cite{Lee:2013iua}. 
In this work, we shall extend such an analysis for $m_{\chi}> 10^4$ GeV as mentioned before. We consider both muon track events and cascade events induced by neutrinos in IceCube observatory. The DM annihilation channels $\chi\chi\rightarrow\tau^{+}\tau^{-}$,
$W^{+}W^{-}$, and $\nu\bar{\nu}$ will be analyzed. Besides analyzing these signature in IceCube, we also study the sensitivity of KM3NeT observatory to the same signature.   
The KM3NeT Observatory \cite{KM3NeT} is a multi-cubic-kilometer scale deep sea neutrino telescope to be built in the Mediterranean Sea. KM3NeT will act as IceCube's counterpart on the Northern hemisphere.
Because of its several cubic kilometers instrumental volume, KM3NeT will be the largest and most sensitive high energy neutrino detector. 
The sensitivities to DM annihilation cross section  $\left\langle \sigma\upsilon\right\rangle $ and DM-proton scattering cross section $\sigma_{\chi p}$ are expected to be enhanced significantly by KM3NeT.

This paper is organized as follows. In Sec. II, we discuss DM capture and annihilation rates inside the Earth and the resulting neutrino flux. We note that the neutrino flux in this case depend on both DM annihilation cross section  $\left\langle \sigma\upsilon\right\rangle $ and DM-proton scattering cross section $\sigma_{\chi p}$.   In Sec. III, we discuss 
the track and shower event rates resulting from DM annihilation in the Earth core. The background event rates from atmospheric neutrino flux are also calculated. In Section IV, we compare signature and background event rates and obtain sensitivities of neutrino telescopes to DM-proton scattering cross section. We present those sensitivities in both isospin symmetric and isospin violating \cite{Kurylov:2003ra,Feng:2011vu} cases, respectively.  We present the summary and conclusion in Section V.

\section{Dark matter annihilation in the Earth core}

\subsection{DM capture and annihilation rates in the Earth core}

The neutrino differential flux $\Phi_{\nu_{i}}$ from $\chi\chi\rightarrow f\bar{f}$
can be expressed as
\begin{equation}
\frac{d\Phi_{\nu_{i}}}{dE_{\nu_{i}}}=P_{\nu,\textrm{surv}}(E_{\nu})\frac{\Gamma_{A}}{4\pi R_{\oplus}^{2}}\sum_{f}B_{f}\left(\frac{dN_{\nu_{i}}}{dE_{\nu_{i}}}\right)_{f}\label{eq:neutrino_flux}
\end{equation}
where $R_{\oplus}$ is the Earth radius, $P_{\nu,\textrm{surv}}$
is the neutrino survival probability from the source to the detector,
$B_{f}$ is the branching ratio of the annihilation channel $\chi\chi\rightarrow f\bar{f}$, $dN_{\nu_{i}}/dE_{\nu_{i}} $ is the energy spectrum of $\nu_i$ produced  per DM annihilation  in the Earth's core, and $\Gamma_{A}$
is the DM annihilation rate in the Earth. To compute $dN_{\nu}/dE_{\nu}$,
we employed \texttt{WimpSim} \cite{Blennow:2007tw} with a total of
50,000 Monte-Carlo generated events.

The annihilation rate, $\Gamma_{A}$, can be obtained by solving the
DM evolution equation in the Earth core \cite{Olive:1986kw,Srednicki:1986vj},
\begin{equation}
\dot{N}=C_{C}-C_{A}N^{2}-C_{E}N\label{eq:evolution_eq}
\end{equation}
where $N$ is the DM number density in the Earth core, $C_{C}$
is the capture rate, and $C_{E}$ is the evaporation rate. The evaporation
rate only relevant when $m_{\chi}\lesssim5\textrm{ GeV}$ \cite{Gould:1987ir,Krauss:1985aaa,Nauenberg:1986em} while
a more refined calculation found typically $m_{\chi}\lesssim3.3\textrm{ GeV}$
\cite{Griest:1986yu}, which are much lower than our interested mass scale.
Thus $C_{E}$ can be ignored in this work. The detail discussion and
derivation to the evolution equation Eq.~\eqref{eq:evolution_eq}
can be found in Ref.~\cite{Gould:1987ir,Krauss:1985aaa,Nauenberg:1986em,Griest:1986yu,Jungman:1995df}.
Solving Eq.~\eqref{eq:evolution_eq} thus gives the annihilation
rate
\begin{equation}
\Gamma_{A}=\frac{C_{C}}{2}\tanh^{2}\left(\frac{t}{\tau_{\oplus}}\right), \label{eq:annihilation_rate}
\end{equation}
where $\tau_{\oplus}$ is the time scale when the DM capture and
annihilation in the Earth core reaches the equilibrium state. Taking $t\approx10^{17}\textrm{ s}$  the lifetime of the solar
system, we have
\begin{equation}
\frac{t}{\tau_{\oplus}}\approx1.9\times10^{4}\left(\frac{C_{C}}{\textrm{s}^{-1}}\right)^{1/2}\left(\frac{\left\langle \sigma\upsilon\right\rangle }{\textrm{cm}^{3}\textrm{ s}^{-1}}\right)^{1/2}\left(\frac{m_{\chi}}{10\textrm{ GeV}}\right)^{3/4},\label{eq:t_over_tE}
\end{equation}
where $\left\langle \sigma\upsilon\right\rangle $ is the DM annihilation
cross section, $m_{\chi}$ is the DM mass, and $C_{C}$ is the
DM capture rate which can be expressed as \cite{Jungman:1995df}
\begin{equation}
C_{C}\propto\left(\frac{\rho_{0}}{0.3\textrm{ GeV cm}^{-3}}\right)\left(\frac{270\textrm{ km s}^{-1}}{\bar{\upsilon}}\right)\left(\frac{\textrm{GeV}}{m_{\chi}}\right)\left(\frac{\sigma_{\chi p}^{0}}{\textrm{pb}}\right)\sum_{i}F_{A_{i}}(m_{\chi}),\label{eq:capture_rate}
\end{equation}
where $\rho_{0}$ is the local DM density, $\bar{\upsilon}$ is the
DM velocity dispersion, $\sigma_{\chi p}^{0}$ is the DM-nucleon
cross section by assuming isospin conservation and $F_{A_{i}}(m_{\chi})$
is the product of various factors for element $A_{i}$ including the
mass fraction, chemical element distribution, 
kinematic suppression, form-factor and reduced mass.

\subsection{Isospin violation effects to bounds set by direct and indirect searches}

\begin{figure}
\begin{centering}
\includegraphics[width=0.49\textwidth]{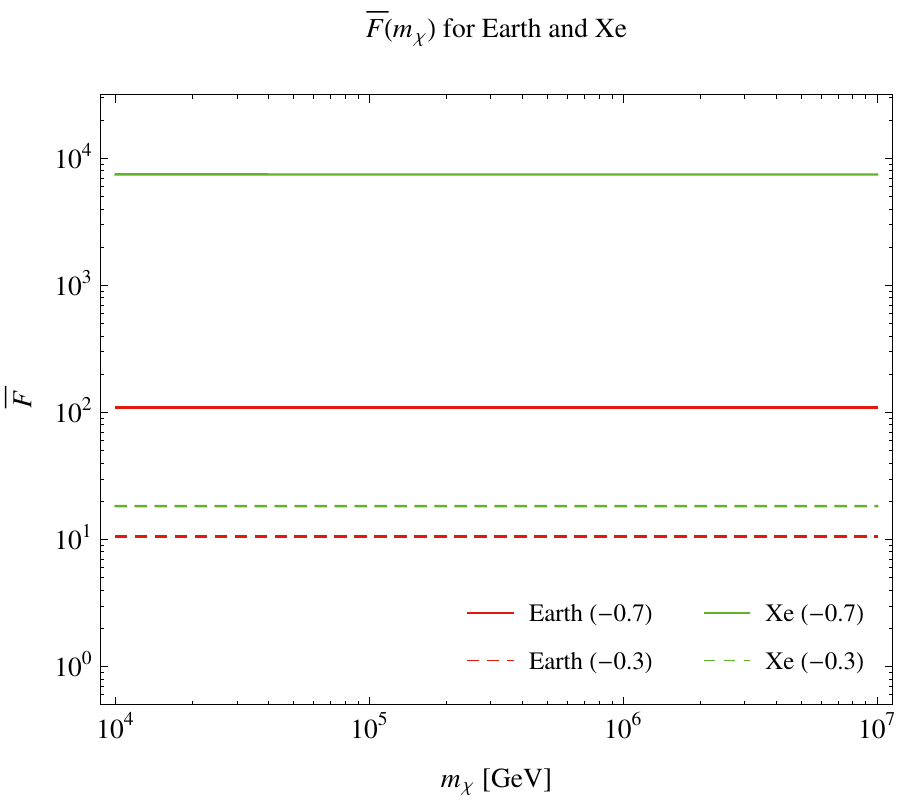}
\par\end{centering}

\caption{\label{fig:Fz} Isospin violation effect for different targets. For xenon target, $\overline{F}$ reduces to $F_Z$. 
In this case, $F_Z$ is as large as $10^4$ for
$f_{n}/f_{p}=-0.7$. With the Earth as the target, $\overline{F}\equiv \sum_Z f_ZF_Z$ with $f_Z$ the fraction of proton targets originating from chemical elements with the atomic number $Z$.
}
\end{figure}

Recent studies \cite{Kurylov:2003ra,Giuliani:2005my,Chang:2010yk,Feng:2011vu}
suggested that DM-nucleon interactions do not necessarily respect the
isospin symmetry. It has been shown that~\cite{Chang:2010yk,Feng:2011vu,Gao:2011bq}
isospin violation can dramatically change the bound on $\sigma_{\chi p}$
from the current direct search. Therefore isospin violation effect is
also taken into consideration in our analysis.

Given an isotope with atomic number $Z$, atom number $A_{i}$,  and
the reduced mass $\mu_{A_{i}}\equiv m_{\chi}m_{A_{i}}/(m_{\chi}+m_{A_{i}})$ for the isotope and the DM particle,
the usual DM-nucleus cross section with the approximation $m_{p}\approx m_{n}$
can be  written as \cite{Jungman:1995df}
\begin{equation}
\sigma_{\chi A_{i}}=\frac{4\mu_{A_{i}}^{2}}{\pi}[Zf_{p}+(A_{i}-Z)f_{n}]^{2}=\frac{\mu_{A_{i}}^{2}}{\mu_{p}^{2}}\left[Z+(A_{i}-Z)\frac{f_{n}}{f_{p}}\right]^{2}\sigma_{\chi p}
\end{equation}
where $f_{p}$ and $f_{n}$ are the effective couplings of DM to protons
and neutrons, respectively. Thus, following
Ref.~\cite{Gao:2011bq}, it is convenient to define the ratio between
$\sigma_{\chi p}$ and $\sigma_{\chi p}^{0}$ where the former is the derived bound on DM-proton cross section with isospin violation while the latter is the derived bound for  isospin symmetric case. For a particular  species of chemical element with atomic number 
$Z$, we have 
\begin{equation}
\frac{\sigma_{\chi p}}{\sigma_{\chi p}^{0}}=\frac{\sum_{i}\eta_{i}\mu_{A_{i}}^{2}A_{i}^{2}}{\sum_{i}\eta_{i}\mu_{A_{i}}^{2}[Z+(A_{i}-Z)f_{n}/f_{p}]^{2}}\equiv F_{Z}\label{eq:isospin_violation}
\end{equation}
where $\eta_{i}$ is the percentage of the isotope $A_{i}$. We note that for a target containing multiple species of chemical elements, the factor $F_Z$ should be modified into $\overline{F}\equiv \sum_Z f_ZF_Z$, where $f_Z$ is the fraction of proton targets originating from elements with the atomic number $Z$. Fig.~\ref{fig:Fz}
shows the numerical values of $\overline{F}$ for different $m_{\chi}$ when
$f_{n}/f_{p}\neq1$. Since $m_{\chi}$ is taken to be larger than $10^4$ GeV, $\overline{F}$ is insensitive to $m_{\chi}$. 

\section{DM signal and atmospheric background events\label{sec:Signal_&_bkg}}

\begin{figure}
\begin{centering}
\includegraphics[width=0.49\textwidth]{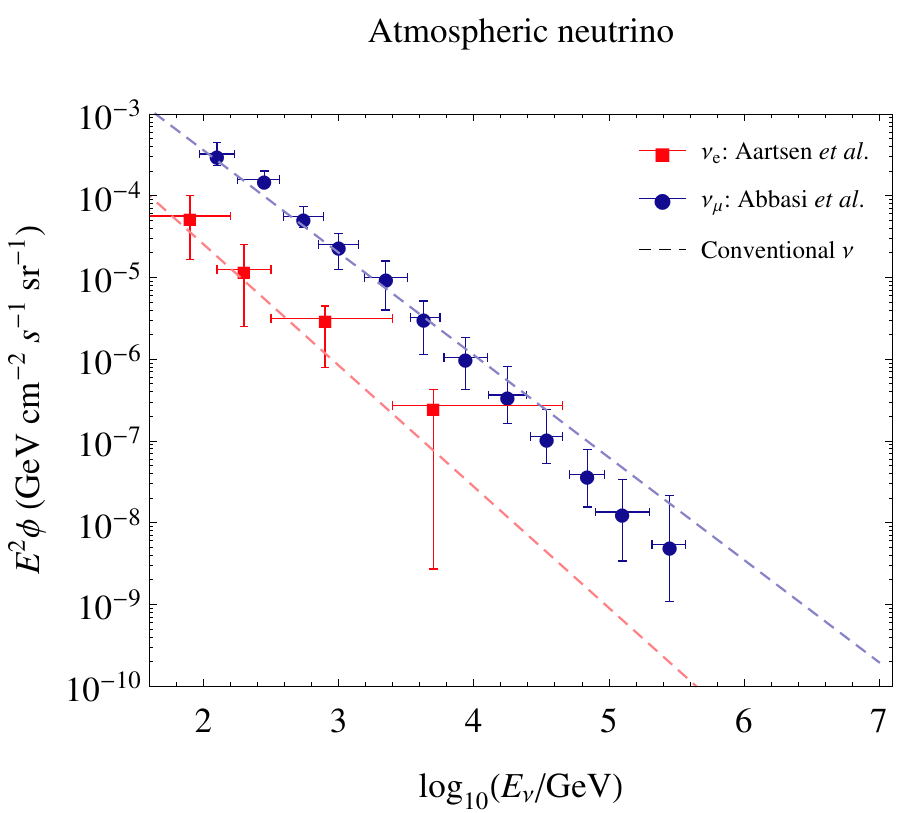}
\par\end{centering}

\caption{\label{fig:atm_nu}The atmospheric $\nu_{e}$ and $\nu_{\mu}$ flux.}
\end{figure}

The neutrino event rate in the detector resulting from DM annihilation in Earth's core is 
\begin{equation}
N_{\nu}=\int_{E_{\textrm{th}}}^{m_{\chi}}\frac{d\Phi_{\nu}}{dE_{\nu}}A_{\nu}(E_{\nu})dE_{\nu}d\Omega\label{eq:nu_event}
\end{equation}
where $E_{\textrm{th}}$ is the detector threshold energy, $d\Phi_{\nu}/dE_{\nu}$
is the neutrino flux from DM annihilation, $A_{\nu}$ is the contained
detector effective area and $\Omega$ is the solid radian. The neutrino
oscillation is strongly suppressed when neutrino carries energy beyond
tens of GeV. Thus the only significant effect to our interested DM
mass region is neutrino attenuation during the propagation from the
production point to the detector. The attenuation effect is included
in the survival probability, $P_{\nu,\textrm{surv}}(E_{\nu})$,
in Eq.~\eqref{eq:neutrino_flux}.

After arriving at the detector, neutrinos are able to produce track or
cascade events through neutral current (N.C.) or charge current (C.C.)
interactions with the medium enclosed by the detector. In this work,
we consider track and cascade events both inside the detector like
IceCube. To compute the event rates in IceCube, the contained effective
areas $A_{\nu}$ for different neutrino flavors in Eq.~\eqref{eq:nu_event}
can be estimated from the effective volume, $V_{\textrm{eff}}$, in
Ref.~\cite{Aartsen:2013jdh} by the following relation:
\begin{equation}
A_{\nu}(E_{\nu})=V_{\textrm{eff}}\frac{N_{A}}{M_{\textrm{ice}}}(n_{p}\sigma_{\nu p}(E_{\nu})+n_{n}\sigma_{\nu n}(E_{\nu}))\label{eq:effective_area}
\end{equation}
where $N_{A}$ is the Avogadro constant, $M_{\textrm{ice}}$ is the
molar mass of ice, $n_{p,n}$ is the number density of proton/neutron
per mole of ice and $\sigma_{\nu p,n}$ is the neutrino-proton/neutron
cross section. Simply swaps the sign $\nu\rightarrow\bar{\nu}$ for
anti-neutrino.

We note that another neutrino telescope KM3NeT located in the
norther-hemisphere is also capable to detect the neutrino signal from DM
annihilation in the Earth. In the present stage, KM3NeT has $\nu_{\mu}$
C.C. effective area published \cite{Katz:2011zza}. Therefore we consider
only track events in KM3NeT.

The atmospheric background event rate is similar to Eq.~\eqref{eq:nu_event},
by replacing $d\Phi_{\nu}/dE_{\nu}$ with atmospheric neutrino flux,
\begin{equation}
N_{\textrm{atm}}=\int_{E_{\textrm{th}}}^{E_{\textrm{max}}}\frac{d\Phi_{\nu}^{\textrm{atm}}}{dE_{\nu}}A_{\nu}(E_{\nu})dE_{\nu}d\Omega.\label{eq:atm_event}
\end{equation}
To facilitate our calculation, the atmospheric neutrino flux $d\Phi_{\nu}^{\textrm{atm}}/dE_{\nu}$ shown in Fig.~\ref{fig:atm_nu}
is taken from Refs.~\cite{Aartsen:2012uu,Honda:2006qj} and extrapolated to $E_{\nu}\simeq 10^{7}$ GeV. We set $E_{\textrm{max}}=m_{\chi}$ in order
to compare with the DM signal.

\section{Results\label{sec:Results}}

We present the sensitivity as a $2\sigma$ detection significance
in 5 years, calculated with the convention,
\begin{equation}
\frac{s}{\sqrt{s+b}}=2.0\label{eq:convention_eq}
\end{equation}
where $s$ is the DM signal, $b$ is the atmospheric background, and
$2.0$ refers to the $2\sigma$ detection significance. The atmospheric
$\nu_{\tau}$ flux is extremely small and can be ignored in our analysis.
Thus we take $\nu_{e}$ and $\nu_{\mu}$ as our major background sources.
The detector threshold energy $E_{\textrm{th}}$ in Eqs.~\eqref{eq:nu_event}
and \eqref{eq:atm_event} is set to be $10^{4}$ GeV in order to suppress
the background events. In the following two subsections, we present
two isospin scenarios for the constraints on $\left\langle \sigma\upsilon\right\rangle $
and $\sigma_{\chi p}$. One is $f_{n}/f_{p}=1$, the isospin symmetry
case, and the other is $f_{n}/f_{p}=-0.7$, the isospin violation
one. Isospin violation scenario is often used to alleviate the inconsistency between the results of different DM direct detection experiments for low $m_{\chi}$. $f_{n}/f_{p}=-0.7$ is the value for which the $\sigma_{\chi p}^{\textrm{SI}}$ sensitivity of a xenon detector is maximally suppressed by isospin violation. Although our study focus on heavy DM accumulated inside the Earth and xenon is very rare among the constituent elements of the Earth, we 
shall see that $f_{n}/f_{p}\sim- 0.7$ leads to most optimistic IceCube sensitivities on both $\left\langle \sigma\upsilon\right\rangle $ and $\sigma_{\chi p}^{\textrm{SI}}$. In the next subsection, we present various $f_{n}/f_{p}$ values
and their impacts to the IceCube sensitivities to the annihilation channel $\chi\chi\rightarrow\tau^{+}\tau^{-}$.

To derive sensitivities to DM-annihilation cross section $\left\langle \sigma\upsilon\right\rangle $,
we make use of the $\sigma_{\chi p}$ from the extrapolation of the
LUX bound \cite{Akerib:2013tjd} to $m_{\chi}>10$ TeV. We find that the total rate $R$
measured by the direct search is given by $R\propto\sigma_{\chi p}\rho_{0}/m_{\chi}m_{A_i}$
for $m_{\chi}\gg m_{A_i}$~\cite{Jungman:1995df} with $\rho_{0}$ the local DM density
and $m_{A_i}$ the mass of the target with $i$ the index for isotopes. Thus $\sigma_{\chi p}\propto m_{\chi}m_{A_i}R/\rho_{0}$
and it is reasonable to extrapolate LUX bound linearly in the
mass scale when $m_{\chi}>10$ TeV.

\subsection{IceCube sensitivities\label{sub:IceCube_sensitivities}}

\begin{figure}
\begin{centering}
\includegraphics[width=0.49\textwidth]{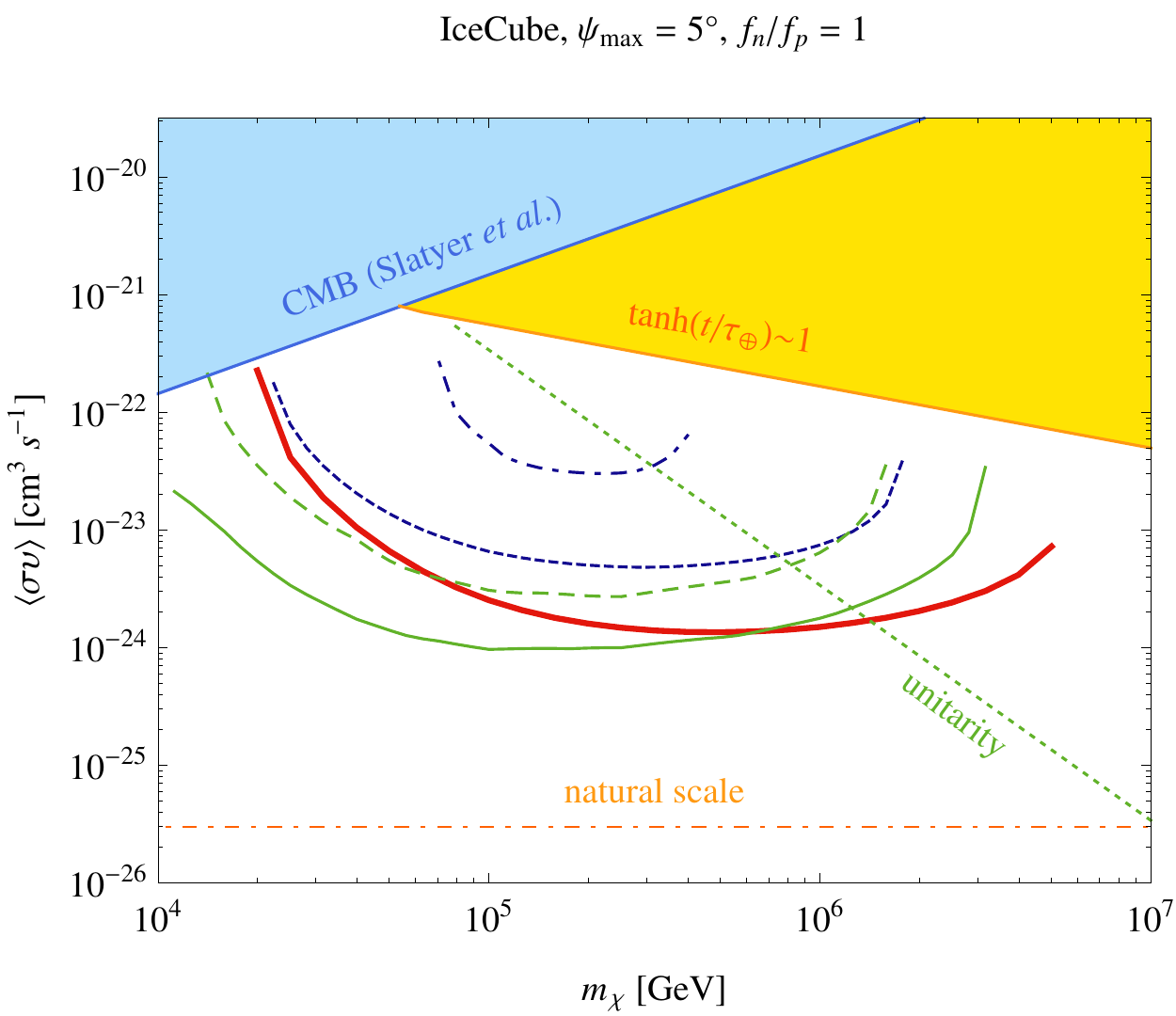}\includegraphics[width=0.49\textwidth]{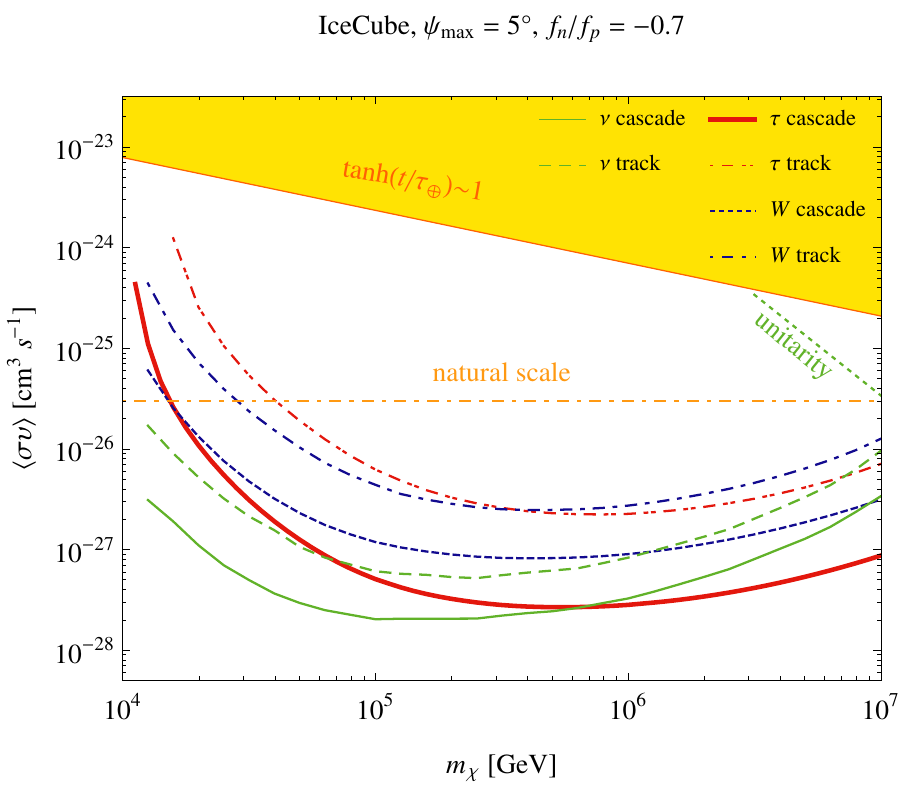}
\par\end{centering}

\caption{\label{fig:IceCube_SV}The IceCube 5-year sensitivity at $2\sigma$
to $\left\langle \sigma\upsilon\right\rangle $ for $\chi\chi\rightarrow\tau^{+}\tau^{-}$,
$W^{+}W^{-}$, and $\nu\bar{\nu}$ annihilation channels with track
and cascade events with $\psi_{\textrm{max}}=5^{\circ}$.
The isospin symmetry case, $f_{n}/f_{p}=1$, is presented on the left panel,
and the isospin violation case, $f_{n}/f_{p}=-0.7$, is presented
on the right panel. The yellow-shaded region is the parameter space for
the equilibrium state and the blue-shaded region is the constraint
from CMB \cite{Slatyer:2009yq}. }

\end{figure}

In Fig.~\ref{fig:IceCube_SV} we present the IceCube sensitivities
to $\left\langle \sigma\upsilon\right\rangle $ of $\chi\chi\rightarrow\tau^{+}\tau^{-}$,
$W^{+}W^{-}$, and $\nu\bar{\nu}$ annihilation channels in the Earth
core with both track and cascade events. For the $\chi\chi\rightarrow\nu\bar{\nu}$ production mode, we assume
equal-flavor distribution ($1/3$ for each flavor). In the left panel where $f_n=f_p$,
the IceCube sensitivities to $\chi\chi\to \tau^{+}\tau^{-}$ and
$\chi\chi\to W^{+}W^{-}$ annihilation channels with track events are only available in
a narrow DM mass range. For most of the DM mass range considered here, the estimated sensitivities are either disfavored 
by the CMB constraint
or reach into the equilibrium region where the $2\sigma$ sensitivity cannot be achieved.
The raising tails for all sensitivities are due to the neutrino
attenuation in the high energy such that larger $\left\langle \sigma\upsilon\right\rangle$
is required to generate sufficient number of events.

For $m_{\chi}\gtrsim10^{6}\textrm{ GeV}$, it is seen that IceCube is more sensitive to $\chi\chi\rightarrow\tau^{+}\tau^{-}$ than to $\chi\chi\rightarrow\nu\bar{\nu}$ for
cascade events. This can be understood by the fact that the neutrino spectrum from
$\chi\chi\rightarrow\nu\bar{\nu}$ is almost like a spike near $m_{\chi}$.
As $m_{\chi}$ becomes larger, neutrinos produced by the annihilation are subject to
more severe energy attenuation. On the other hand, the neutrino spectrum from $\chi\chi\rightarrow\tau^{+}\tau^{-}$
is relatively flat in the whole energy range. The energy attenuation only affects
the higher energy neutrinos.

\begin{figure}
\begin{centering}
\includegraphics[width=0.49\textwidth]{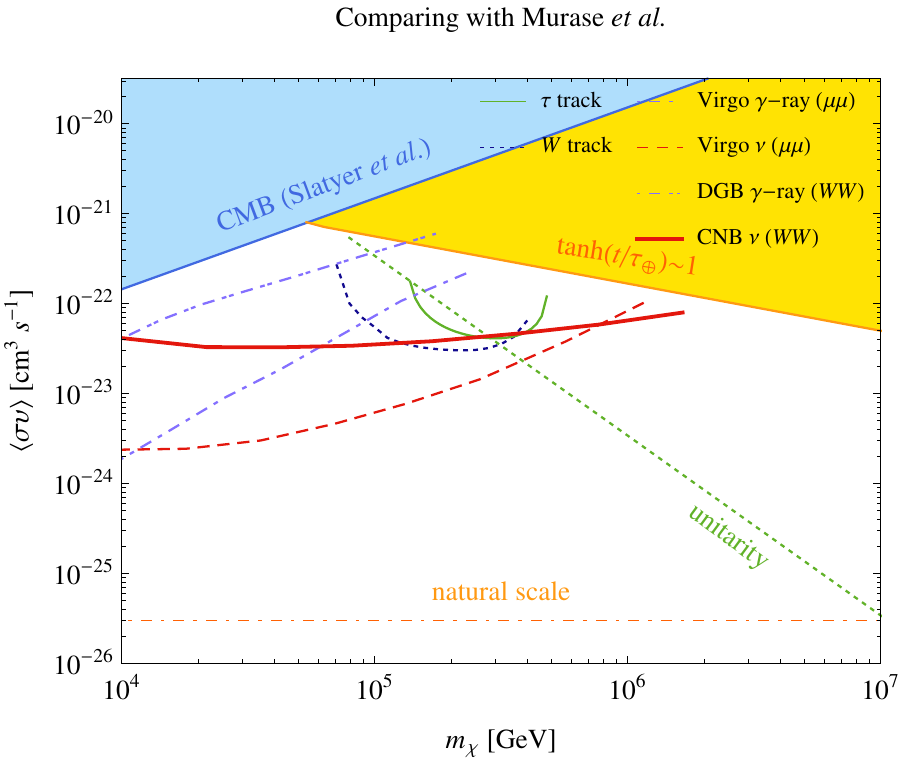}
\par\end{centering}

\caption{\label{fig:MuraseAll} The IceCube 5-year sensitivity at $2\sigma$
to $\left\langle \sigma\upsilon\right\rangle $ for $\chi\chi\rightarrow W^{+}W^{-}$ and
$\tau^{+}\tau^{-}$ annihilation channels for track
events with $\psi_{\textrm{max}}=5^{\circ}$, respectively. The dot-dashed line is the gamma-ray constraint on the $\chi\chi\rightarrow\mu^{+}\mu^{-}$ annihilation cross section in Virgo cluster~\cite{Murase:2013}. The dashed line is the projected full IceCube $2\sigma$ sensitivity in 5 years to $\left\langle \sigma (\chi\chi\rightarrow\mu^{+}\mu^{-})\upsilon\right\rangle $ in Virgo cluster in the presence of substructures with track events~\cite{Murase:2013}. The dot-dot-dashed line is the cascade gamma-ray constraint on $\left\langle \sigma (\chi\chi\rightarrow W^{+}W^{-}) \upsilon\right\rangle$ from diffuse gamma-ray background (DGB)~\cite{Murase:2012}. The thick solid line is the full IceCube sensitivity in 3 years to $\left\langle \sigma (\chi\chi\rightarrow W^{+}W^{-}) \upsilon\right\rangle$ from cosmic neutrino background (CNB) with track events~\cite{Murase:2012}. }
\end{figure}

In the isospin violation scenario, the ratio $f_{n}/f_{p}=-0.7$ could
weaken the LUX bound by four orders of magnitude, i.e., the LUX upper
bound on $\sigma_{\chi p}$ is raised by four orders of magnitude. Taking
a four orders of magnitude enhanced $\sigma_{\chi p}$, the DM capture rate 
is enhanced by two orders of magnitude since the suppression factor
due to isospin violation is around $10^{-2}$ for chemical elements
in the Earth's core. With the DM capture rate enhanced by two orders of magnitude,
the IceCube sensitivities to $\left\langle \sigma\upsilon\right\rangle $
of various annihilation channels can be improved by about four orders
of magnitude by simple scaling observed in Refs.~\cite{Albuquerque:2011ma,Lee:2013iua}. Therefore, the sensitivities could reach below the natural
scale $\left\langle \sigma\upsilon\right\rangle =3\times10^{-26}\textrm{ cm}^{2}\textrm{ s}^{-1}$.

\begin{figure}
\begin{centering}
\includegraphics[width=0.49\textwidth]{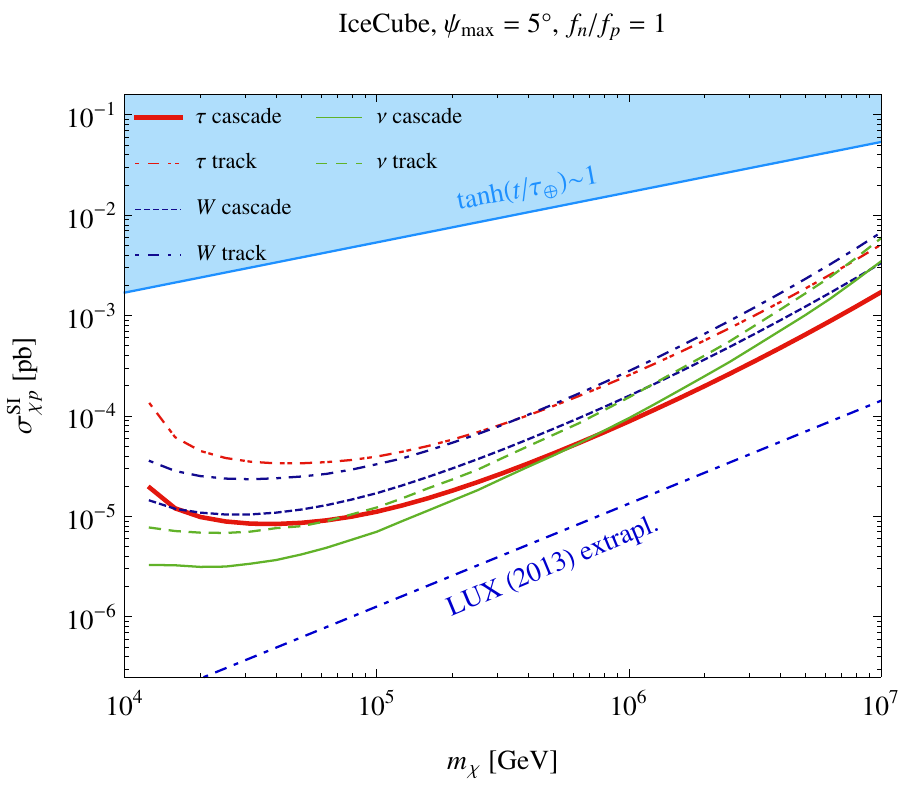}\includegraphics[width=0.49\textwidth]{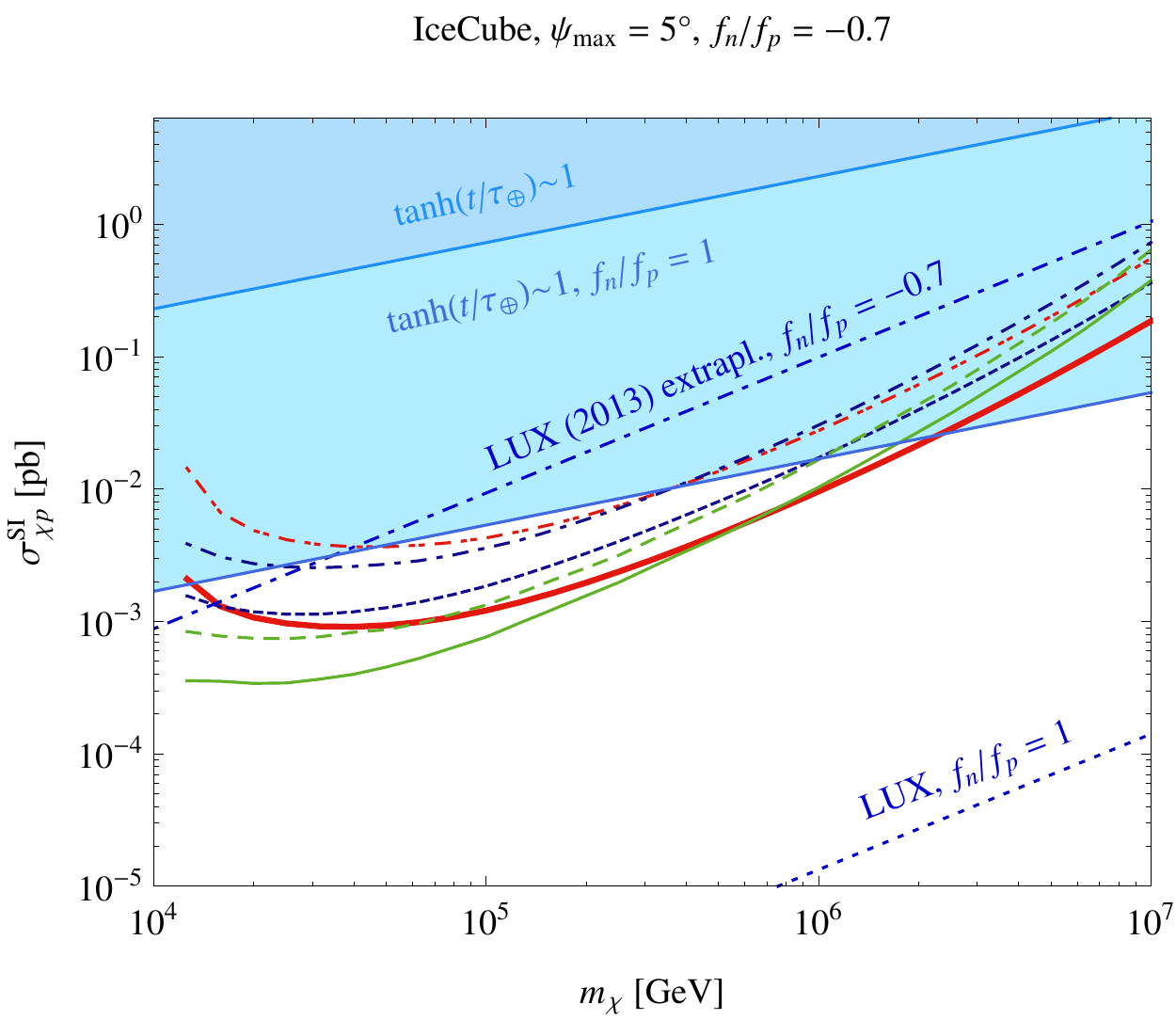}
\par\end{centering}

\caption{\label{fig:IceCube_SI}The IceCube $2\sigma$ sensitivities in 5 years
to $\sigma_{\chi p}^{\textrm{SI}}$ for $\chi\chi\rightarrow\tau^{+}\tau^{-}$,
$W^{+}W^{-}$, and $\nu\bar{\nu}$ annihilation channels with both
track and cascade events with $\psi_{\textrm{max}}=5^{\circ}$. The
isospin symmetry case, $f_{n}/f_{p}=1$, is presented on the left,
and the isospin violation case, $f_{n}/f_{p}=-0.7$, is presented
on the right. The blue-shaded region is the parameter space for the
equilibrium state and the light-blue-shaded region on the right panel
refers to the equilibrium-state parameter space for the isospin symmetry
case as a comparison. An extrapolation of current LUX limit has been
shown on the figures.}
\end{figure}

For DM annihilation, a general upper bound on $\left\langle \sigma\upsilon\right\rangle$ is set by unitarity condition \cite{Griest:1990,Hui:2001,Beacom:2007}. 
The DM annihilation   
cross section is assumed to be $s$-wave dominated in the low-velocity limit. Hence it can be shown that~\cite{Beacom:2007}
\begin{equation}
\left\langle \sigma\upsilon\right\rangle\leq\dfrac{4\pi}{m^{2}_{\chi}\upsilon}\simeq 1.5\times 10^{-13}\ \dfrac{\rm{cm}^{3}}{\rm{s}}
\left( \dfrac{\rm{GeV}}{m_{\chi}} \right) ^{2} \left( \dfrac{300\ \rm{km}/\rm{s}}{\upsilon_{\rm{rms}}} \right) .
\end{equation}    
This unitarity bound with $\upsilon_{\rm{rms}} \simeq 13\ \rm{km}\ \rm{s^{-1}}$(escape velocity from the Earth) is also shown in Fig.~\ref{fig:IceCube_SV}. 
The unitarity bound can be evaded for non-point-like DM particles \cite{Griest:1990,Hui:2001,Murase:2012}.

Galaxy clusters (GCs) are the largest gravitationally bound objects in the universe and their masses can be as large as $10^{15}$ times that of the Sun's ($10^{15} M_{\odot}$) \cite{Voit:2005,Diaferio:2008}. Many galaxies (typically $\sim$ 50 - 1000) collect into GCs, but their masses consist of mainly  dark matter. Thus GCs are the largest DM reservoirs in the universe and can be the ideal sources to look for DM annihilation signatures. With DM particles assumed to annihilate into $\mu^{+}\mu^{-}$ pairs, the predicted full IceCube $2\sigma$ sensitivity in 5 years to $\left\langle \sigma\upsilon\right\rangle $ for Virgo cluster in the presence of substructures with track events is derived in Ref.~\cite{Murase:2013}. We present this sensitivity in Fig.~\ref{fig:MuraseAll} and we can see that it is better than our expected IceCube 5-year sensitivity at $2\sigma$ to $\left\langle \sigma (\chi\chi\rightarrow\tau^{+}\tau^{-}) \upsilon\right\rangle $ with $\nu_{\mu}$ track events. One of the reasons is because only 18$\%$ of $\tau$ decay to $\nu_{\mu}$. However, if we consider isospin violation scenario, 
our expected IceCube sensitivity with  $f_{n}/f_{p}=-0.7$ will be much better than that for Virgo cluster. Except for neutrinos,
DM annihilation in GCs can also produce a high luminosity in gamma rays. In Ref.~\cite{Murase:2013}, the authors also estimate gamma-ray constraints taking into account electromagnetic cascades caused by pair production on the cosmic photon backgrounds, from the flux upper limits derived by Fermi-LAT observations of GCs~\cite{Pinzke:2011,Ackermann:2010}. We show in Fig.~\ref{fig:MuraseAll} the gamma-ray constraint on the $\chi\chi\rightarrow\mu^{+}\mu^{-}$ annihilation cross section for Virgo cluster taken from Ref.~\cite{Murase:2013}. We can see that this constraint is weaker than our expected IceCube 5-year sensitivity at $2\sigma$ to $\left\langle \sigma (\chi\chi\rightarrow\tau^{+}\tau^{-}) \upsilon\right\rangle $ for $m_{\chi} \gtrsim 10^{5}$ GeV.

The diffuse gamma-ray background (DGB) was measured by Fermi Large Area Telescope (Fermi-LAT) above 200 MeV in 2010~\cite{Abdo:2010}. Radio-loud active galactic nuclei (AGN) including blazars~\cite{Abazajian:2011}, star-forming and star-burst galaxies~\cite{Fields:2010,Loeb:2006}, and heavy DM are the possible sources. 
In Ref.~\cite{Murase:2012}, the authors derive cascade gamma-ray constraints on the annihilation cross section of heavy DM by requiring the calculated cascade gamma-ray flux not exceeding the measured DGB data at any individual energy bin by more than a given significance~\cite{Fermi:2010,Ackermann:2012}. We present the cascade gamma-ray constraint on $\left\langle \sigma (\chi\chi\rightarrow W^{+}W^{-}) \upsilon\right\rangle$ for DGB taken from Ref.~\cite{Murase:2012} in Fig.~\ref{fig:MuraseAll}. We note that this constraint is weaker than our predicted IceCube 5-year sensitivity at $2\sigma$ to $\left\langle \sigma (\chi\chi\rightarrow W^{+}W^{-}) \upsilon\right\rangle $. On the other hand, for demonstrating the power of neutrino observations, we also show in Fig.~\ref{fig:MuraseAll} the predicted full IceCube sensitivity in 3 years to $\left\langle \sigma (\chi\chi\rightarrow W^{+}W^{-}) \upsilon\right\rangle$ for cosmic neutrino background (CNB) with track events taken from Ref.~\cite{Murase:2012}. It is slightly less sensitive compared to our expected IceCube 5-year sensitivity at $2\sigma$ to $\left\langle \sigma (\chi\chi\rightarrow W^{+}W^{-}) \upsilon\right\rangle $ at $\sim 10^{5}$ GeV, while both sensitivities do not reach to the unitarity bound for $m_{\chi} \gtrsim 3\times 10^{5}$ GeV.

Fig.~\ref{fig:IceCube_SI} shows the IceCube sensitivities to spin-independent
cross section $\sigma_{\chi p}^{\textrm{SI}}$ by analyzing track
and cascade events from $\chi\chi\rightarrow\tau^{+}\tau^{-}$, $W^{+}W^{-}$,
and $\nu\bar{\nu}$ annihilation channels in the Earth core. The threshold
energy $E_{\textrm{th}}$ is the same as before and we take $\left\langle \sigma\upsilon\right\rangle =3\times10^{-26}\textrm{ cm}^{2}\textrm{ s}^{-1}$
as our input. Precisely speaking, the sensitivity to $\chi\chi\rightarrow\nu\bar{\nu}$
channel is the highest when $m_{\chi}\lesssim10^{6}\textrm{ GeV}$
and $\chi\chi\rightarrow\tau^{+}\tau^{-}$ after. However, the sensitivities
to different channels can be taken as comparable since the differences
between them are not significant.

When isospin is a good symmetry, the IceCube sensitivity is not as good as 
the constraint from the LUX extrapolation. However, with
$f_{n}/f_{p}=-0.7$, the capture rate 
is reduced to 1\% of the isospin symmetric value. Therefore one requires
100 times larger $\sigma_{\chi p}^{\textrm{SI}}$ to reach the same
detection significance. However, the ratio $f_{n}/f_{p}=-0.7$ makes
a more dramatic impact to the DM direct search using xenon as the
target. The DM scattering cross section with xenon is reduced by four orders of magnitude. Hence the indirect
search by IceCube could provide better constraint on $\sigma_{\chi p}^{\textrm{SI}}$
than the direct search in such a case.

\subsection{KM3NeT sensitivities}

\begin{figure}
\begin{centering}
\includegraphics[width=0.49\textwidth]{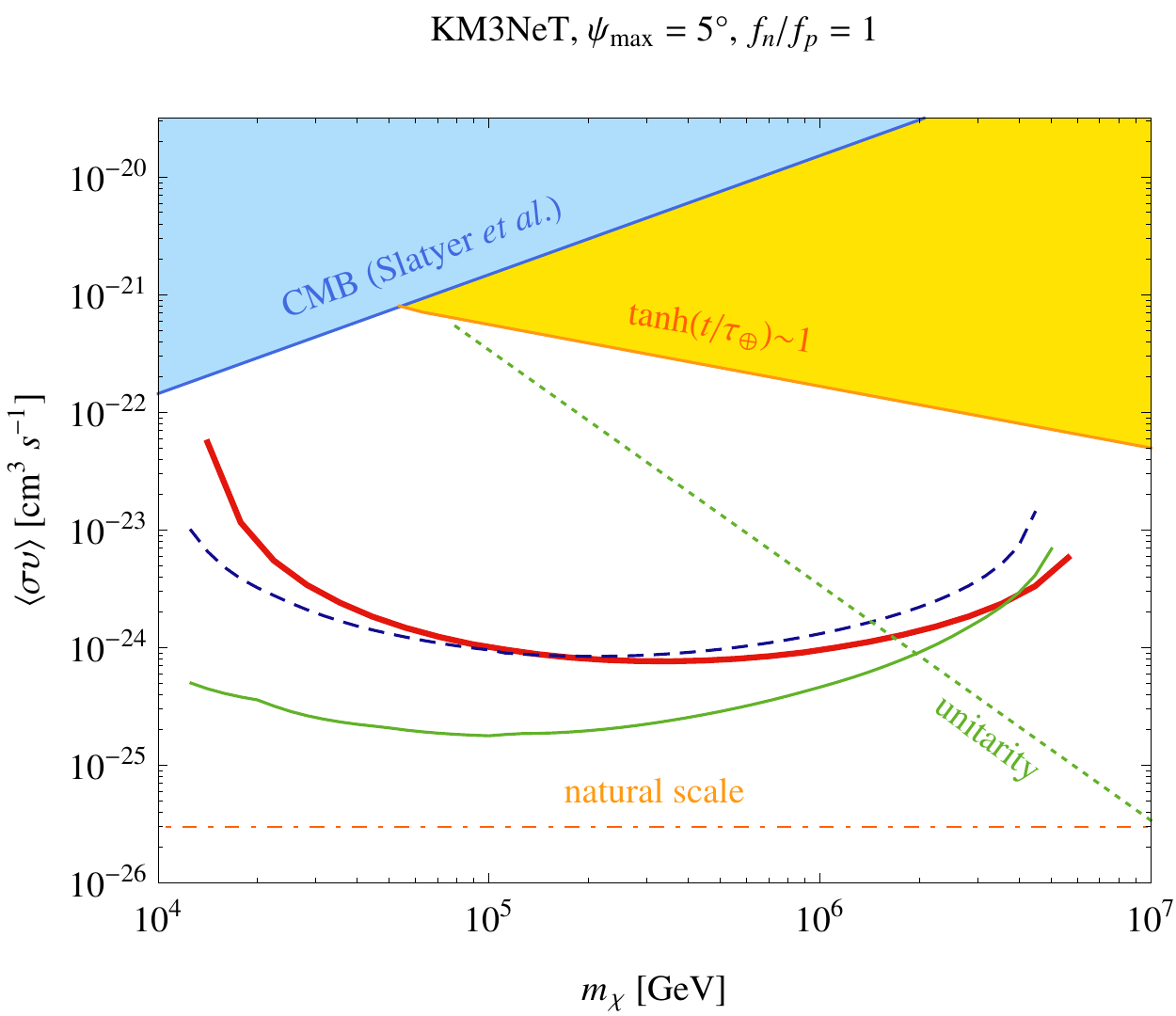}\includegraphics[width=0.49\textwidth]{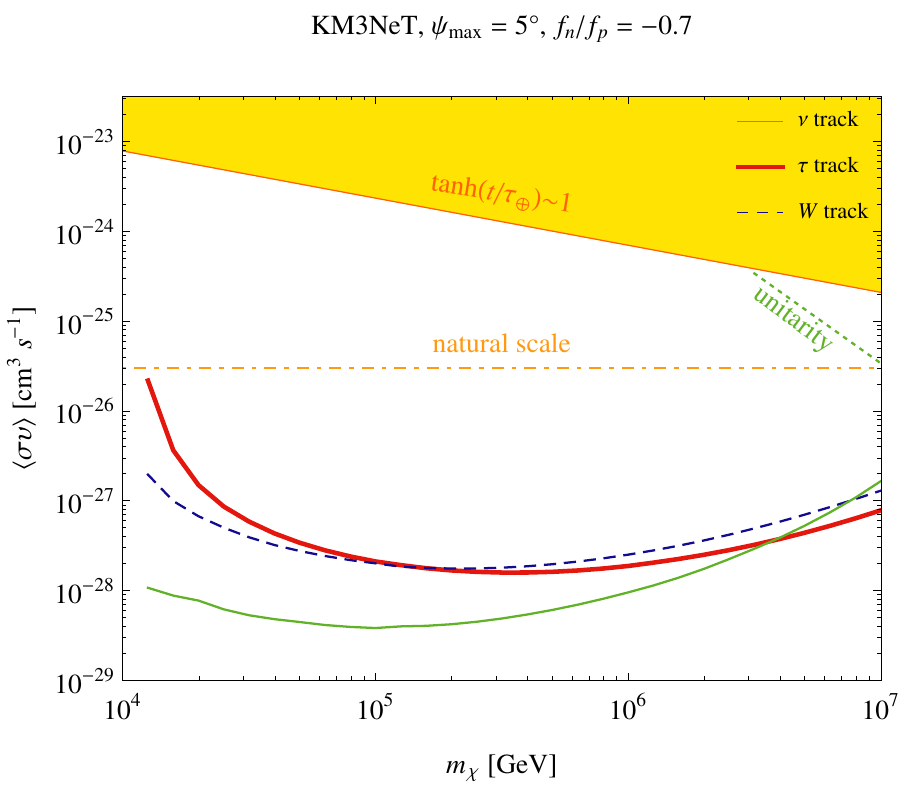}
\par\end{centering}

\caption{\label{fig:KM3SV}The KM3NeT $2\sigma$ sensitivities in 5 years to
$\left\langle \sigma\upsilon\right\rangle $ for $\chi\chi\rightarrow\tau^{+}\tau^{-}$,
$W^{+}W^{-}$, and $\nu\bar{\nu}$ annihilation channels with track
events only and $\psi_{\textrm{max}}=5^{\circ}$. The isospin symmetry
case, $f_{n}/f_{p}=1$, is presented on the left panel, and the isospin
violation case, $f_{n}/f_{p}=-0.7$, is presented on the right panel. The
yellow-shaded region is the parameter space for the equilibrium state
and the blue-shaded region is the constraint from CMB \cite{Slatyer:2009yq}.}
\end{figure}
\begin{figure}
\begin{centering}
\includegraphics[width=0.49\textwidth]{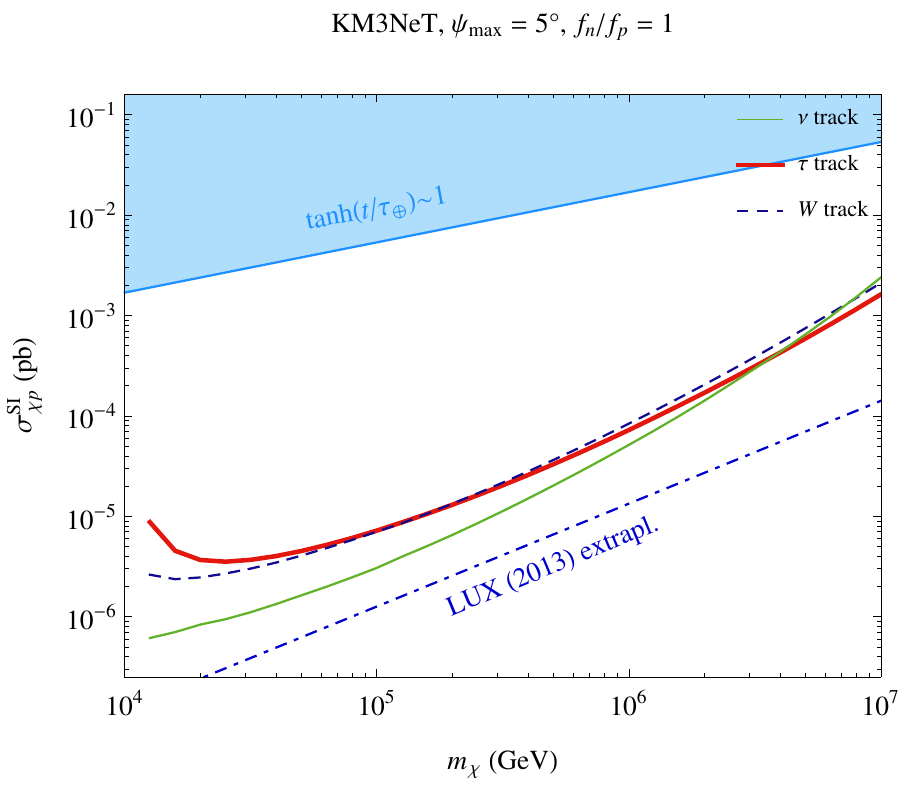}\includegraphics[width=0.49\textwidth]{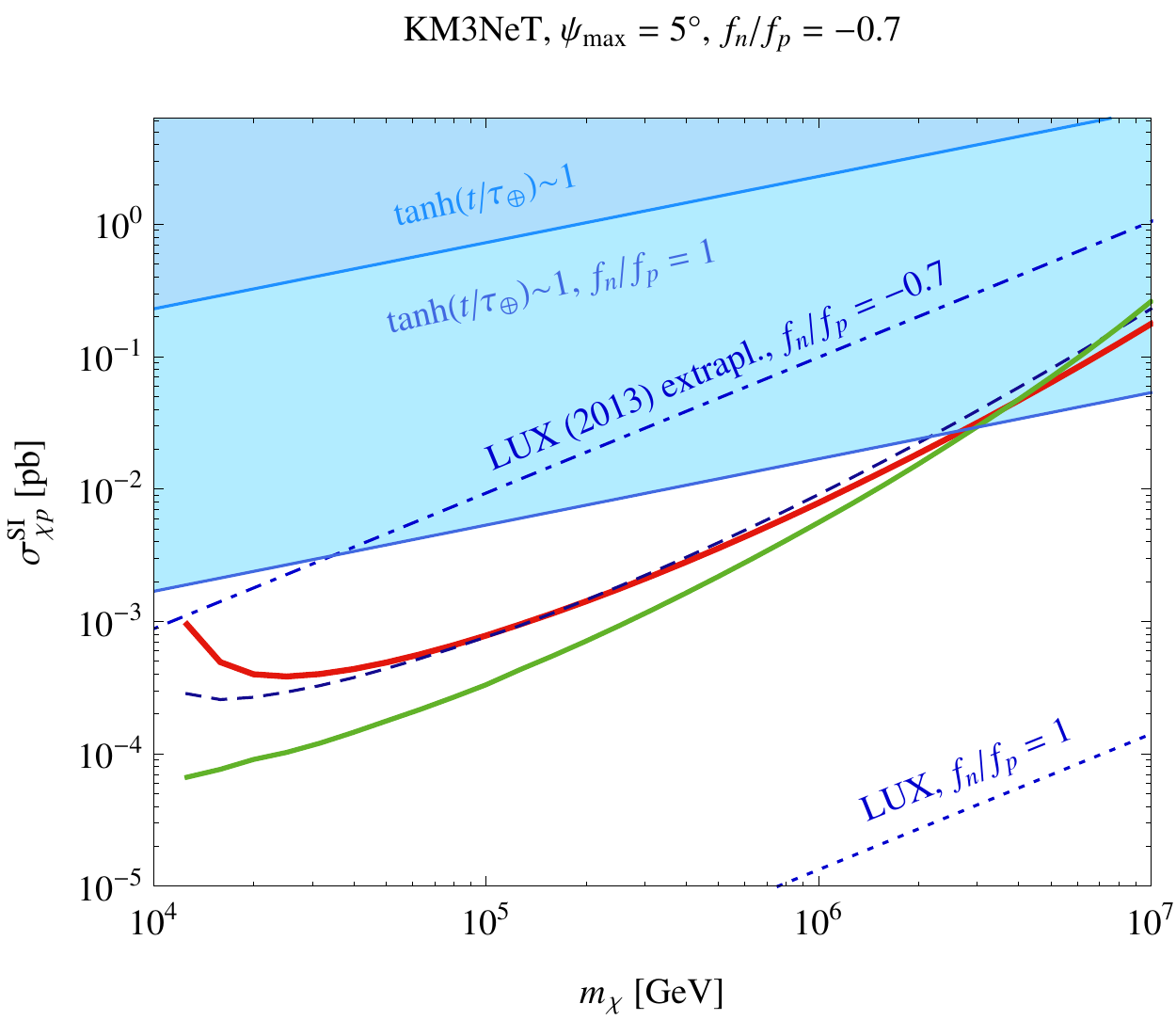}
\par\end{centering}

\caption{\label{fig:KM3SI}The KM3NeT $2\sigma$ sensitivities in 5 years to
$\sigma_{\chi p}^{\textrm{SI}}$ for $\chi\chi\rightarrow\tau^{+}\tau^{-}$,
$W^{+}W^{-}$, and $\nu\bar{\nu}$ annihilation channels for track
events only with $\psi_{\textrm{max}}=5^{\circ}$. The isospin symmetry
case, $f_{n}/f_{p}=1$, is presented on the left panel, and the isospin
violation case, $f_{n}/f_{p}=-0.7$, is presented on the right panel. The
blue-shaded region is the parameter space for the equilibrium state
and the light-blue-shaded region on the right panel refers to the
equilibrium-state parameter space in the isospin symmetry case.}
\end{figure}

Besides IceCube, the neutrino telescope KM3NeT located in the northern-hemisphere
can also reach to a promising sensitivity in the near future \cite{Biagi:2012mg}.
Therefore it is worthwhile to comment on the performance of KM3NeT.
Since KM3NeT only published $\nu_{\mu}$ charge-current effective
area in the present stage, we shall only analyze track events.

The results are shown in Fig.~\ref{fig:KM3SV} and \ref{fig:KM3SI}
with parameters chosen to be the same as those for computing the IceCube
sensitivities. The sensitivities of KM3NeT are almost
an order of magnitude better than those of IceCube, since its $\nu_{\mu}$
charge-current effective area is about one order of magnitude larger than that of IceCube.

\subsection{Sensitivities  with different $f_n$ to $f_p$ ratios}

\begin{figure}
\begin{centering}
\includegraphics[width=0.49\textwidth]{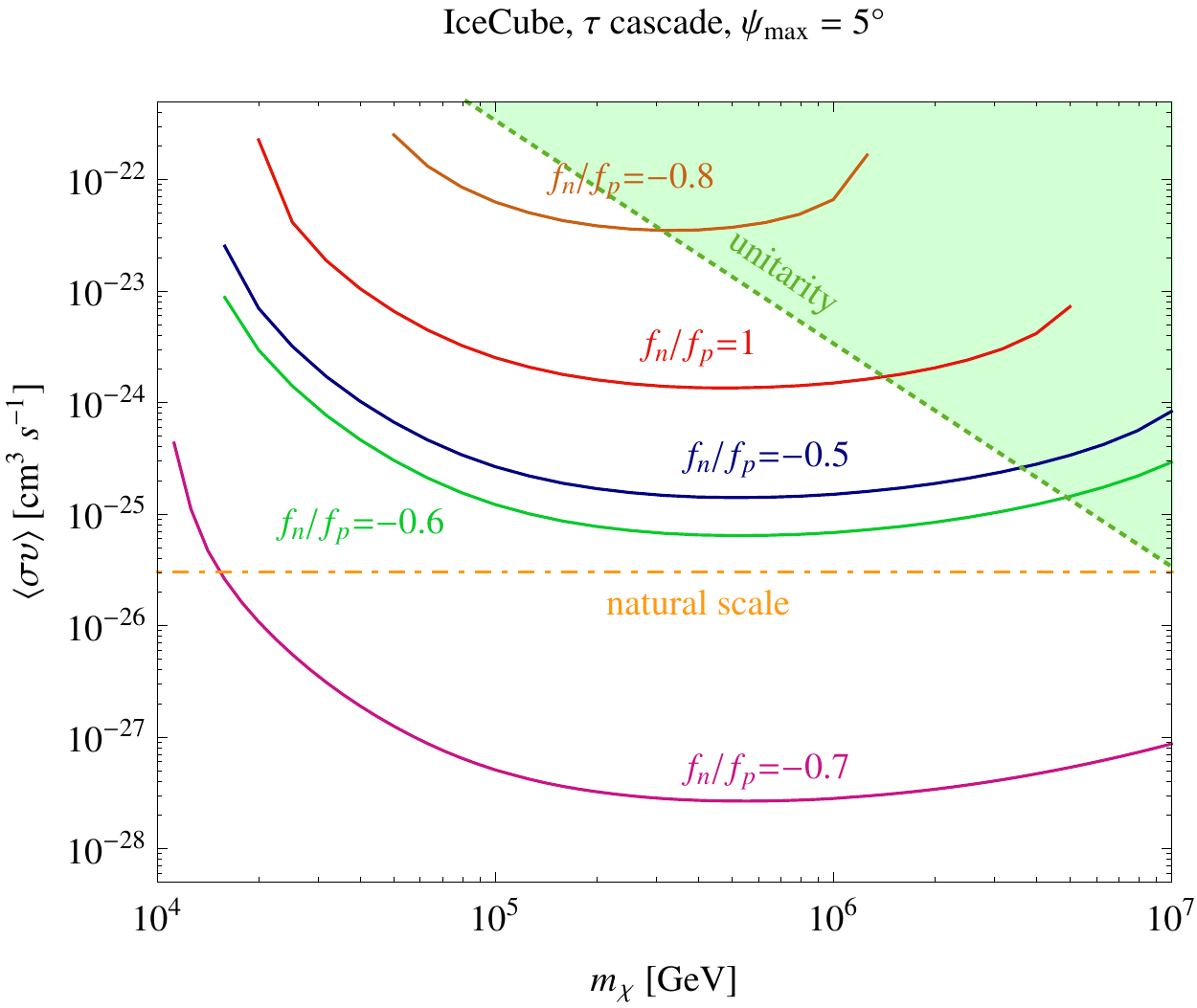}\includegraphics[width=0.49\textwidth]{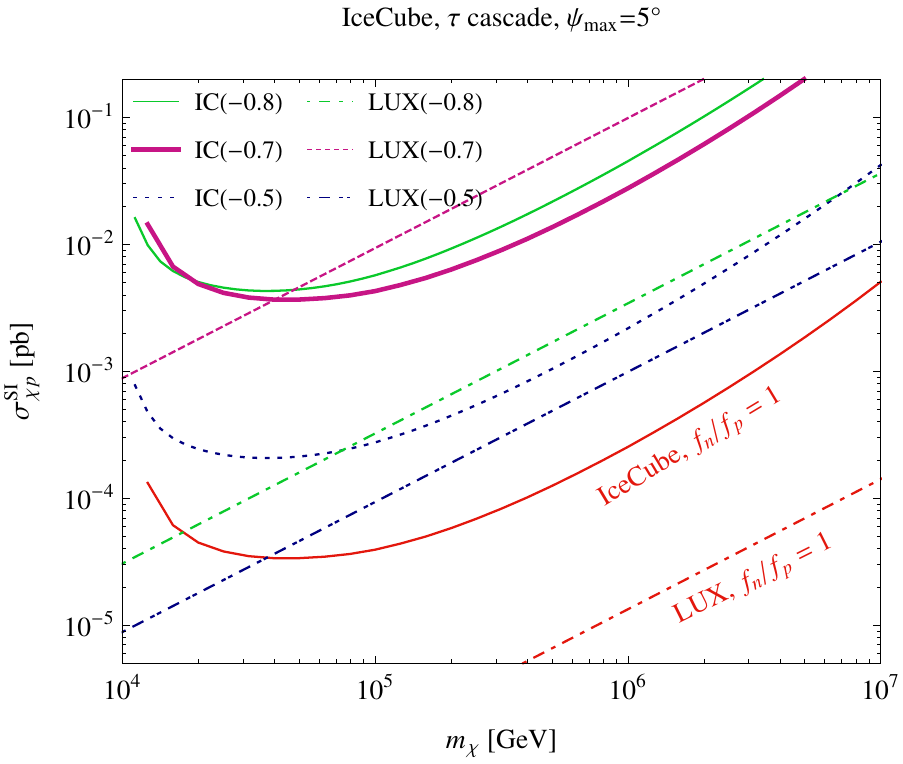}
\par\end{centering}

\caption{\label{fig:IC_SV_Multi_IV}The IceCube 5-year sensitivity at $2\sigma$
to $\left\langle \sigma\upsilon\right\rangle $ on the left panel and $\sigma_{\chi p}^{\textrm{SI}}$
on the right panel for $\chi\chi\rightarrow\tau^{+}\tau^{-}$ annihilation
channels with cascade events for different degrees of isospin violation.
We take the re-derived $\sigma_{\chi p}^{\textrm{SI}}$ from LUX with isospin violation taken into consideration. }
\end{figure}

In the previous subsections, we have presented IceCube and KM3NeT  sensitivities to $\left\langle \sigma\upsilon\right\rangle $
and $\sigma_{\chi p}^{\textrm{SI}}$ for $f_{n}/f_{p}=1$ and $-0.7$.
To be thorough, it is worth discussing the effect to DM search with
various $f_{n}/f_{p}$ values. For simplicity, we shall focus
on the $\chi\chi\rightarrow\tau^{+}\tau^{-}$ cascade events in IceCube.

In the left panel of Fig.~\ref{fig:IC_SV_Multi_IV}, we present IceCube sensitivities to $\left\langle \sigma\upsilon\right\rangle$
with $f_{n}/f_{p}\in[-0.8,1]$. We take the re-derived
$\sigma_{\chi p}^{\textrm{SI}}$ from LUX using Eq.~\eqref{eq:isospin_violation}
which quantifies the isospin violation effect. Isospin
violation not only leads to the suppression of DM capture rate by the Earth 
but also weaken the $\sigma_{\chi p}^{\textrm{SI}}$ bound from LUX. The
overall effect is beneficial to the DM indirect search for $f_{n}/f_{p}$
in a certain range. As shown in Fig.~\ref{fig:IC_SV_Multi_IV}, the IceCube sensitivity to
$\left\langle \sigma\upsilon\right\rangle $ improves as $f_{n}/f_{p}\to -0.7$ from the above. However,  when $f_{n}/f_{p}$ is smaller than
$-0.7$, the sensitivity to $\left\langle \sigma\upsilon\right\rangle $ becomes even worse than that in the isospin symmetry case.

In the right panel of Fig.~\ref{fig:IC_SV_Multi_IV}, we present IceCube sensitivities to $\sigma_{\chi p}^{\textrm{SI}}$
with $f_{n}/f_{p}\in[-0.8,1]$ by taking $\left\langle \sigma\upsilon\right\rangle =3\times10^{-26}\textrm{ cm}^{3}\textrm{ s}^{-1}$
as our input. With isospin symmetry violated, the DM capture rate is suppressed by the factor $\bar{F}$ defined right below Eq.~(\ref{eq:isospin_violation}). Thus to
reach the same detection significance by indirect search, one requires
a larger $\sigma_{\chi p}^{\textrm{SI}}$ to produce enough events.
However, isospin violation also weaken the LUX limit at certain range of $f_{n}/f_{p}$. It turns out the sensitivity to $\sigma_{\chi p}^{\textrm{SI}}$
by IceCube is better than the existing limit by LUX only for $f_{n}/f_{p}$ slightly larger or equal to $-0.7$.
For $f_{n}/f_{p}<-0.7$, the LUX limit becomes stringent again while DM capture
rate still suffers from suppression from isospin violation.

\section{Summary and conclusion}

In this work we have presented the IceCube and KM3NeT sensitivities
to thermal-averaged annihilation cross section $\left\langle \sigma\upsilon\right\rangle $
and DM spin-independent cross section $\sigma_{\chi p}^{\textrm{SI}}$
for heavy DM ($m_{\chi}> 10^4$ GeV) by detecting DM induced neutrino signature from the Earth's core.
To probe the former, we take $\sigma_{\chi p}^{\textrm{SI}}$ from the LUX bound~\cite{Akerib:2013tjd} as the input.
To probe the latter, we take  $\left\langle \sigma\upsilon\right\rangle=3\times 10^{-26} $ cm$^3$s$^{-1}$ as the input.
The IceCube sensitivity to $\left\langle \sigma(\chi\chi\to W^+W^-)\upsilon\right\rangle $ in the present case is slightly better than
its sensitivity to $\left\langle \sigma(\chi\chi\to W^+W^-)\upsilon\right\rangle $ in the case of  detecting cosmic neutrino background~\cite{Murase:2012}. 
On the other hand, the IceCube sensitivity to $\left\langle \sigma(\chi\chi\to \tau^+\tau^-)\upsilon\right\rangle $ in the present case is not as good as 
its sensitivity to $\left\langle \sigma(\chi\chi\to \mu^+\mu^-)\upsilon\right\rangle$ 
in the case of  detecting neutrinos from Virgo cluster~\cite{Murase:2013}. Concerning IceCube and 
KM3NeT sensitivities to $\sigma_{\chi p}^{\textrm{SI}}$, we have shown that they are roughly one order of magnitude worse than 
the LUX bound. 

We stress that the above comparison is based upon the assumption of isospin symmetry
in DM-nucleon couplings. We have shown that, like the direct search, the indirect search
is also affected by the isospin violation. The implications of isospin
violation to IceCube and KM3NeT observations have been presented in
Sec.~\ref{sec:Results}. Taking isospin violation effect into account,
the sensitivities of the above neutrino telescopes to both $\left\langle \sigma\upsilon\right\rangle$ and $\sigma_{\chi p}^{\textrm{SI}}$
through detecting the signature of DM annihilation in the Earth's core can be 
significantly improved. As $f_n/f_p\to -0.7$,  the sensitivities to $\left\langle \sigma\upsilon\right\rangle$ can be 
better than the natural scale while the sensitivities to $\sigma_{\chi p}^{\textrm{SI}}$ can be better than the LUX bound..

\acknowledgements

We thank Y.-L. Sming Tsai for
helpful advice in computations. This work is supported by National Science Council
of Taiwan under Grant No.~102-2112-M-009-017. 


\end{document}